\documentclass[12pt]{article}
\usepackage[utf8]{inputenc}
\usepackage{palatino}
\usepackage{indentfirst}
\usepackage[english]{babel}
\usepackage{amssymb, amsmath, amsthm}
\usepackage[inline]{enumitem}
\usepackage{xcolor}
\usepackage{ushort}
\usepackage[nohead,margin=1in]{geometry}
 
\usepackage{setspace}
\usepackage{graphicx}
\usepackage{caption}
\usepackage{subcaption}
\usepackage{float}
\usepackage{natbib}
\usepackage{color}
\usepackage{enumitem}
\usepackage{tikz}
\usepackage[colorlinks = true,
            linkcolor = blue,
            urlcolor  = blue,
            citecolor = blue,
            anchorcolor = blue]{hyperref}
\usetikzlibrary{decorations.pathreplacing}
\tikzset{mybrace/.style={decoration={brace,raise=1.8mm},decorate}}
\newtheoremstyle{example}{}{}{}{}{\color{blue}\bfseries}{.}{ }{}
\DeclareMathOperator\supp{supp}
\newtheorem{theorem}{Theorem}
\usepackage{lipsum}
\newtheorem{proposition}{Proposition}
\newtheorem{corollary}{Corollary}

\newtheorem{lemma}{Lemma}

\theoremstyle{example}
\newtheorem{example}{Example}

\newcommand{\continuation}{??}

\title{Robust Regulation of Firms' Access to Consumer Data\footnote{I am deeply grateful with Piotr Dworczak, Jeff Ely and Alessandro Pavan for their guidance. I would also like to thank Yingni Guo and Harry Pei for their comments. All errors remain my own.}}
\author{Jose Higueras\footnote{Department of Economics, Northwestern University. Email: \href{mailto:josehiguerascorona2025@u.northwestern.edu}{\textcolor{blue}{josehiguerascorona2025@u.northwestern.edu}}}}
\date{\today}

\begin{document}

\maketitle

\begin{abstract}
I study how to regulate firms' access to consumer data when a regulator faces non-Bayesian uncertainty about how firms will exploit the consumer's information to segment the market and set prices. I fully characterize all worst-case optimal policies when the regulator maximizes consumer surplus: the regulator allows a firm to access data only if the firm cannot use the database to identify a \textit{small group} of consumers. 
\end{abstract}

\section{Introduction}
\noindent The growing accessibility of consumer data 
enable firms to obtain unprecedented insights into consumer behavior. The value of such data is diverse, with one notable aspect being its potential to provide information on consumers' willingness to pay. This capability, in turn, may lead to price discrimination, affecting welfare and consumer surplus. The latter raises the question: How should consumer privacy laws regulate firms' access to consumer data when utilizing it for price discrimination? 

To address this question, it is essential to understand how firms utilize consumer data to make pricing decisions. However, recent developments in collecting, transmitting, and analyzing consumer data---along with firms' strong incentives to protect this information---suggest that policymakers and regulators may lack insight into the specific ways data is used for price discrimination. For instance, firms may employ marketing metrics that are inaccessible to regulators, which are crucial for understanding the connection between data and consumers' willingness to pay. Such insights then enable firms to segment the market and set prices accordingly. Additionally, firms can match consumer data with other information, including past purchase decisions---capabilities that regulators do not possess---and use this information for price discrimination. More broadly, the Council of Economic Advisors report on big data and price discrimination \textcolor{blue}{\citep{cea}} states:
\begin{quote}
    ``Although a few studies have tried to detect differential pricing online, current knowledge is
    mainly anecdotal. Companies naturally protect information about pricing strategies for
    competitive reasons, and perhaps also for fear of a customer backlash.'' 
\end{quote}

In this paper, I explore this question by analyzing a scenario in which a regulator concerned about consumer surplus has the authority to control which data a firm can access, yet faces non-Bayesian uncertainty regarding how firms will exploit this consumer information to segment the market and set prices. More formally, I represent this uncertainty by assuming that the regulator lacks knowledge of the correlation structure between the data and consumers' willingness to pay. 

I consider a model in which a monopolist (he) sells a good to a unit mass of consumers. Each consumer is privately informed about their valuation for the good, which takes finitely many values. The monopolist can acquire additional information about consumers through what I call a \textit{database}. A database specifies the correlation between consumers' valuations and a \textit{covariate}, which assigns a label from a finite set to each consumer. This covariate may capture characteristics such as income, gender, or race. The database is valuable because it allows the monopolist to segment the market and engage in third-degree price discrimination.

In contrast, a regulator (she)---aiming to maximize consumer surplus---implements a policy that controls which covariates the monopolist may access. She knows the marginal distributions of valuations and labels, but not their joint distribution; that is, she lacks access to the underlying database. To address this uncertainty, she evaluates each covariate based on its \textit{worst-case} consumer surplus across all joint distributions consistent with the marginals. A policy is \textit{worst-case optimal} if every covariate authorized for access by the regulator yields the maximum worst-case consumer surplus.

My main result, Theorem~\ref{theorem:characterization-WC}, fully characterizes the set of covariates that attain the maximum worst-case consumer surplus, enabling the identification of all worst-case optimal policies available to the regulator. The result follows from a simple observation that applies to any covariate: the joint distribution where labels and valuations are independent is always consistent with the marginals. This scenario reverts to uniform pricing, resulting in the consumer surplus obtained with the uniform monopoly price. Therefore, this consumer surplus is an upper bound on the worst-case consumer surplus. Moreover, a trivial covariate that assigns the same label to all consumers---which resembles the case where the monopolist has no access to information about consumers---always leads to the consumer surplus obtained with uniform pricing. Thus, a covariate yields the maximum worst-case consumer surplus if and only if it achieves a consumer surplus weakly greater than the one obtained with uniform pricing for all plausible joint distributions.

Theorem~\ref{theorem:characterization-WC} provides a simple, sufficient, and necessary condition for a covariate to always achieve a consumer surplus weakly greater than that obtained through uniform pricing. This condition establishes that the proportion of consumers assigned to any label must be higher than a uniform lower bound, which I analytically characterize. In other words, a policy is worst-case optimal if and only if any covariate authorized for access by the regulator does not allow the monopolist to identify a \textit{small} group of consumers, where in the context of this paper consumers belong to the same group if they share the same label and the group is classified as small if the proportion of consumers that belong to it falls below the uniform lower bound from Theorem~\ref{theorem:characterization-WC}. This result can provide an economic rationale for why specific consumer privacy laws restrict access to data that allows firms to identify small or minority groups of consumers. For example, the Equal Credit Opportunity Act (ECOA) prohibits credit companies from using data that discloses consumers' race, ethnicity, or religion when assessing their creditworthiness.

After identifying all the worst-case optimal policies available to the regulator, in the same spirit as the one proposed by \citet{borgers2017no}, I ask the following question: From the set of covariates that attain the maximum worst-case consumer surplus which ones are undominated? That is, for which covariates there is no alternative covariate that never yields lower consumer surplus and sometimes higher consumer surplus.

Theorem~\ref{theorem:characterization-UD} identifies the covariates that yield the maximum worst-case consumer surplus and are also undominated. These covariates not only consistently achieve a consumer surplus weakly higher than that obtained through uniform pricing but can also yield a consumer surplus strictly superior to that attained via the uniform monopoly price for certain feasible joint distributions of valuations and labels. Theorem~\ref{theorem:characterization-WC} establishes both the sufficient and necessary conditions for a covariate to exhibit this property. In addition to satisfying the uniform lower bound set by Theorem~\ref{theorem:characterization-WC}, the proportion of consumers assigned to one of the labels is below an upper bound, which I also characterize. 

Finally, Theorem~\ref{theorem:characterization-WC} and~\ref{theorem:characterization-UD} also generalize to the case where the regulator's objective is to maximize a weighted average of consumer and producer surplus, where the weight assigned to consumer surplus is weakly higher than the weight assigned to producer surplus. 

The remainder of this paper is organized as follows. Section~\ref{sec:literature} reviews the related literature. Section~\ref{sec:model} introduces the model. Section~\ref{sec:results} shows the main results of this paper. Section~\ref{sec:extensions} extends the model and results to alternative objectives for the regulator, and Section~\ref{sec:conclusions} presents concluding remarks.

\subsection{Related Literature}\label{sec:literature}
\noindent This paper contributes to the literature on data markets recently surveyed by \citet{bergemann2019markets}. Within this literature, I am mainly motivated by the insights of \citet{bergemann2015limits}. In this work, the authors characterize all combinations of consumer and producer surplus that can emerge from third-degree price discrimination. This characterization shows that market outcomes can be pretty heterogeneous when the monopolist has additional information about the market beyond the prior distribution. This paper then asks: If a regulator prioritizes consumer surplus, what additional information about the market should she permit the monopolist to access? 

\citet{10.1093/restud/rdac033}, \citet{pram2021disclosure}, \citet{vaidyadisclosure} and \citet{argenziano2021data} also analyze the regulation of data markets. However, my paper differs from these studies in two aspects. First, prior studies approach regulation from a positive perspective, analyzing the welfare effects of granting consumers control over their data. In contrast, my paper adopts a normative approach. Specifically, it explores a different question: If a regulator can enact any policy governing firms' access to data, what form should such policy take? Second, in prior studies, every player has complete knowledge of the relevant environment. In contrast, in my setting, the regulator possesses non-Bayesian uncertainty concerning the correlation between data and willingness to pay.

Another relevant strand of the literature examines mechanism and information design in monopoly settings under worst-case or worst-regret objectives. \citet{du2018robust} examines an auction for selling a common-value good, finding that the minimum revenue across all information that bidders might hold about the good's value converges to the full surplus as the number of bidders increases. \citet{carroll2017robustness} explores a setting where a monopolist sells multiple goods to a buyer. The monopolist knows the marginal distribution of the buyer's valuation for each good but is uncertain about the joint distribution. Concurrently, \citet{che2021robustly} and \citet{deb2021multi} also analyze scenarios in which a monopolist sells multiple goods to a buyer. In \citet{che2021robustly}, the monopolist knows only that the distribution of the buyer’s valuations lies within a certain set, without assigning a prior over it. In \citet{deb2021multi}, the monopolist faces non-Bayesian uncertainty about the information the buyer may acquire regarding her valuations.

In all the setups above, it is the monopolist who faces non-Bayesian uncertainty. In contrast, my paper centers on a regulator who faces non-Bayesian uncertainty about the information available to the monopolist. Related to this idea, \citet{guo2019robust} considers a regulator that has non-Bayesian uncertainty about the demand function and cost function faced by the monopolist.

Lastly, as the comment made by \citet{borgers2017no} on the work by \citet{chung2007foundations} and the work by \citet{dworczak2022preparing}, this paper explores the dominance order among policies that achieve the maximum worst-case payoff.

\section{Model}\label{sec:model}
\noindent A monopolist (he) sells a good to a unit mass of consumers, each of whom demands one unit. The monopolist faces constant marginal costs of production that I normalize to zero. The consumers privately observe their \textit{willingness to pay} (\textit{valuation}) for the good, which can take $K$ possible values, $V \equiv \{v_1,...,v_k,...,v_K\}$, with $v_k \in \mathbb{R}_+$; without loss of generality assume: $0<v_1<\dots<v_K$.

A \textit{market} $x$ is a distribution over the $K$ valuations, with the set of all markets being:
\[
\Delta(V) \equiv \left\{x \in \mathbb{R}^K_{+} \; \Bigg \lvert \; \sum_{k=1}^K x_k=1\right\},
\]
where $x_k$ is the proportion of consumers who have valuation $v_k$. Furthermore, I denote by $\bar{x}_k$ the proportion of consumers who have a valuation greater or equal to $v_k$. That is, $\bar{x}_k=\sum_{j=k}^Kx_j$. Additionally, I say that the price $v_k$ is \textit{optimal} for market $x$ if the expected revenue from price $v_k$ satisfies $v_k \bar{x}_k \geq v_i \bar{x}_i$ for all $i \in \{1,\dots, K\}$.

Throughout the analysis, I fix and denote by $x^* \in \Delta(V)$ the market faced by the monopolist. Let $v_{i^*}$ be the highest price among all optimal prices for market $x^*$:
\[
v_{i^*} = \max \left\{\arg \max_{v_k \in V} \; v_k \bar{x}^*_k\right\}.
\]

I will refer to $v_{i^*}$ as the \textit{uniform monopoly price}. Given this uniform monopoly price, the corresponding consumer surplus is given by:
\[
u^* = \sum_{j=i^*}^K (v_j-v_{i^*})x^*_j.
\]

To avoid trivialities in the analysis, assume that $u^*>0$\footnote{As I will show later in Lemma~\ref{lemma:WC=F1}, a covariate yields the maximum worst-case consumer surplus if and only if it achieves a consumer surplus weakly greater than $u^*$ for all plausible conditional distributions of valuations given labels. If $u^*=0$, then any covariate yields the maximum worst-case consumer surplus. To avoid this trivial case, I assume that $u^*>0$.}. This implies that there is a valuation $v_k$ higher than the uniform monopoly price $v_{i^*}$, and the proportion of consumer that have valuation $v_k$ is strictly positive. That is, $x^*_k>0$.

\subsection{Data}
\noindent
Formally, a \textit{database} is a joint distribution $d \in \Delta(V \times S)$ over the set of valuations $V$ and a finite set of labels $S$, where $d_{k,s}$ denotes the share of consumers with valuation $v_k$ and label $s$. Let $f \in \Delta(S)$ denote the marginal distribution over labels. Without loss of generality, assume full support: $f_s > 0$ for all $s \in S$. I refer to the pair $(S,f)$ as a \textit{covariate}. 

Since the specific labels in $S$ play no intrinsic role in the analysis, we normalize by letting $S \equiv \{1,2,\dots,n\}$ and identify each covariate solely by its marginal distribution $f$. The set of all possible covariates---marginal distributions over finite label sets with full support---is given by:
\[
F \equiv \bigcup_{n \in \mathbb{N}} \left\{ f \in \mathbb{R}^n_{++} \;\middle|\; \sum_{s=1}^n f_s = 1 \right\}.
\]

A special case arises when $n = 1$, yielding a unique trivial covariate with singleton support, denoted by $f^{\emptyset}$. A database with covariate $f^{\emptyset}$ corresponds to a setting in which no information is available---that is, all consumers share the same label.

\subsection{Value of Data}
\noindent
Given a database $d$ with covariate $f$, define $\sigma(\cdot \mid s) \in \Delta(V)$ as the conditional distribution over valuations given label $s$. Specifically, $\sigma(k \mid s) = \frac{d_{k,s}}{f_s}$ represents the share of consumers with valuation $v_k$ conditional on label $s$. I refer to $\sigma(\cdot \mid s)$ as \textit{segment} $s$, and I denote the full collection of these conditional distributions by $\sigma \equiv \left\{\sigma(\cdot \mid s)\right\}_{s=1}^n$, calling this collection a \textit{segmentation}.

Note that since $d$ must be consistent with the marginals $x^*$ and $f$, the segmentation decomposes the aggregate market:
\[
\sum_{s=1}^n f_s \sigma(k \mid s) = \sum_{s=1}^n d_{k,s} = x^*_k, \quad \text{for all } k \in \{1,\dots,K\}.
\]

If the monopolist has access to covariate $f$, he also has access to the corresponding database $d$. That is, the monopolist knows the joint distribution of valuations and labels. As a result, he will set profit-maximizing prices within each segment. In segment $s$, the chosen price is $v_{i^*_s}$, where:
\begin{equation} \label{eq:monopoly-price}
v_{i^*_s} = \max \left\{ \arg\max_{v_k \in V} \; v_k \sum_{j=k}^K \sigma(j \mid s) \right\}.
\end{equation}
If multiple prices yield the same profit, ties are broken against the consumer.

Given these prices, the associated consumer surplus is:
\[
\underline{u}(\sigma,f) = \sum_{s=1}^n f_s \sum_{j=i^*_s}^K (v_j - v_{i^*_s}) \sigma(j \mid s).
\]

\subsection{Regulation}
\noindent
A regulator (she), concerned with consumer surplus, has the authority to control which covariate the monopolist may access. Formally, the regulator selects a policy $\mathcal{F} \subseteq F$, and the monopolist can access one covariate from this set.

Recall that if the monopolist has access to $f$, he also has access to the corresponding database $d$. In contrast, the regulator is assumed to know only the market distribution $x^*$ and faces non-Bayesian uncertainty about the underlying joint distribution $d$. To address this, she evaluates each covariate based on its \textit{worst-case consumer surplus}, defined as the minimum attainable surplus across all joint distributions consistent with $x^*$ and $f$. For any $f$, define the set of \textit{feasible segmentations} as:
\[
\Sigma(f) \equiv \left\{ \sigma(\cdot \mid s) \in \Delta(V) \; \forall s \in \{1,\dots,n\} \;\middle|\; \sum_{s=1}^n f_s \sigma(\cdot \mid s) = x^* \right\}.
\]

Then, the regulator evaluates $f$ by:
\[
\inf_{\sigma \in \Sigma(f)} \underline{u}(\sigma,f).
\]

A policy $\mathcal{F}$ is said to be \textit{worst-case optimal} if every $f \in \mathcal{F}$ yields the maximum worst-case consumer surplus. That is, $\mathcal{F} \subseteq WC$, where:
\[
WC \equiv \arg\max_{f \in F} \inf_{\sigma \in \Sigma(f)} \underline{u}(\sigma,f).
\]

The following Lemma gives a simple expression for $WC$ and shows that it is non-empty.

\begin{lemma}\label{lemma:WC=F1}
Let $F_1=\left\{f \in F \; \lvert \; \underline{u}(\sigma,f) \geq u^* \; \forall \; \sigma \in \Sigma(f)\right\}$ be the set of covariates that yield a consumer surplus greater or equal to $u^*$ for all feasible segmentations. Then, $f^{\emptyset} \in F_1$ and $WC \equiv F_1$.
\end{lemma}
\begin{proof}
    For any $f \in F$, $\sigma(\cdot \lvert s)=x^*$, for all $s$, belongs to $\Sigma(f)$. In other words, the joint distribution where labels and valuations are independent is always consistent with the marginals. In this case, charging the uniform monopoly price in every segment is optimal; therefore, consumer surplus equals $u^*$. Moreover, $u^*$ is an upper bound on the worst-case consumer surplus. Additionally, for $f^{\emptyset} \in F$, the only feasible consumer surplus is $u^*$. Hence, $f^{\emptyset} \in WC$, and $f \in WC$ if and only if $\underline{u}(\sigma,f) \geq u^*$ for all $\sigma \in \Sigma(f)$.
\end{proof}

\section{Results}\label{sec:results}
\noindent I start by presenting the main result of this paper. This result fully characterizes the set of covariates that yield the maximum worst-case consumer surplus. 

\begin{theorem}\label{theorem:characterization-WC}
    $f$ yields a consumer surplus greater or equal to $u^*$ for all $\sigma \in \Sigma(f)$ if and only if $f_s > \underline{\lambda}$ for every label $s$, where: 
    \[
    \underline{\lambda} =  \min_{v_j \leq v_{i^*}} \; \frac{\max_{v_i>v_{i^*}} \; v_i \Bar{x}^*_{i}}{v_j}+1-\Bar{x}^*_j.
    \]
\end{theorem}
\begin{proof}[If direction of Theorem~\ref{theorem:characterization-WC}]
     Suppose that $f_s > \underline{\lambda}$ for all $s$. Towards a contradiction, assume that there is a $\sigma \in \Sigma(f)$ such that $\underline{u}(\sigma,f)<u^*$. Then, a segment $s$ must exist where the monopolist sets a price higher than $v_{i^*}$. Denote such a price by $v_k$. Since $v_k$ maximizes the monopolist's profits in segment $s$, for all $v_i < v_k$ it follows that:
     
    \begin{align}\label{eq:lowbound-sigma-WC}
    v_k\sum_{j=k}^{K}\sigma(j \lvert s) - v_i \sum_{j=i}^{K}\sigma(j \lvert s)={v_i}\sum_{j=k}^{K}\sigma(j \lvert s)+\sum_{j=1}^{i-1}\sigma(j \lvert s) -1&\geq 0.
    \end{align}
    
    Furthermore, since $\sigma \in \Sigma(f)$, $\sigma(\cdot \lvert s)f_s \leq x^*$. Then, if I multiply both sides of inequality~\ref{eq:lowbound-sigma-WC} by $f_s$ I get that:
    
    \begin{align}\label{eq:upbound-f-WC}
    \frac{v_k}{v_i}\bar{x}^*_k + 1-\bar{x}^*_i&= \frac{v_k}{v_i}\sum_{j=k}^{K}x^*_j + \sum_{j=1}^{i-1}x^*_j\geq\left[\frac{v_k}{v_i}\sum_{j=k}^{K}\sigma(j \lvert s)+\sum_{j=1}^{i-1}\sigma(j \lvert s)\right]f_s\geq f_s.
    \end{align}
    
    Inequality~\ref{eq:upbound-f-WC} holds for some $v_k > v_{i^*}$ and all $v_i \leq v_{i^*}$, therefore:
    \begin{equation}
        f_s \leq  \min_{v_i \leq v_{i^*}}\frac{\max_{v_k > v_{i^*}} v_k\bar{x}^*_{k}}{v_i}+ 1-\bar{x}^*_i=\underline{\lambda}, 
    \end{equation}
    which is a contradiction.
\end{proof}

In other words, $f$ yields the maximum worst-case consumer surplus if and only if the proportion of consumers assigned to any label is above the uniform lower bound $\underline{\lambda}$. One potential interpretation of this result is that the regulator allows access to a covariate only if the monopolist cannot exploit it to identify a small group of consumers. In the context of this paper, consumers are considered part of the same group if they share a label, and a group is defined as small if the proportion of consumers within it falls below $\underline{\lambda}$.

The proof of the only if part of Theorem~\ref{theorem:characterization-WC} is provided in Appendix~\ref{sec:proofs-WC}. It builds on the algorithm proposed by \citet{bergemann2015limits}, which segments $x^*$ using a ``greedy'' procedure that generates segments corresponding to extreme markets. An extreme market is defined as a distribution of valuations in which the monopolist is indifferent between charging any price in the support.

Specifically, I show that the total mass of consumers assigned by the algorithm to extreme markets containing valuations strictly greater than the uniform monopoly price equals $\underline{\lambda}$. I then use this result to construct a feasible segmentation that yields a consumer surplus strictly lower than $u^*$ whenever $f_s \leq \underline{\lambda}$ for some $s$. The following example illustrates how such a segmentation is constructed.

\begin{example} The set of valuations is given by $V =\{1, 2, 3, 4\}$. The market faced by the monopolist is represented by the distribution $x^* = \left(\frac{2}{5}, \frac{1}{10}, \frac{2}{5}, \frac{1}{10}\right)$, where $x^*_1 = \frac{2}{5}$, $x^*_2 = \frac{1}{10}$, $x^*_3 = \frac{2}{5}$, and $x^*_4 = \frac{1}{10}$. The uniform monopoly price is 3, which yields a consumer surplus of $u^* = \frac{1}{10}$. In this case, it is straightforward to verify that the uniform lower bound is $\underline{\lambda}=\frac{2}{5}$.

Using the greedy algorithm proposed by \citet{bergemann2015limits}\footnote{For a formal definition of the greedy algorithm, see the proof of the only if direction of Theorem~\ref{theorem:characterization-WC} in Appendix~\ref{sec:proofs-WC}.}, I obtain the following segmentation of $x^*$:
\begin{table}[H]
\centering
\caption{Extremal segmentation for $x^*$}
\label{tab1}
\begin{tabular}{@{}cccccc@{}@{}}
\hline
Extreme markets
& \multicolumn{1}{c}{$v=1$}
& \multicolumn{1}{c}{$v=2$}
& \multicolumn{1}{c}{$v=3$}
& \multicolumn{1}{c}{$v=4$}
& Mass of consumers \\
\hline
$x^{\{1,2,3,4\}}$ & $\frac{1}{2}$ & $\frac{1}{6}$ & $\frac{1}{12}$ & $\frac{1}{4}$ & $\frac{2}{5}$ \\
$x^{\{1,2,3\}}$ & $\frac{1}{2}$ & $\frac{1}{6}$ & $\frac{1}{3}$ & $0$ & $\frac{1}{5}$ \\ 
$x^{\{1,3\}}$ & $\frac{2}{3}$ &  $0$ &  $\frac{1}{3}$ & 0 & $\frac{3}{20}$\\
$x^{\{3\}}$ & $0$ &  $0$ & 1 & $0$ & $\frac{1}{4}$\\
\hline
$x^*$ & $\frac{2}{5}$ & $\frac{1}{10}$ & $\frac{2}{5}$ & $\frac{1}{10}$ & $1$
\end{tabular}
\end{table}

The segmentation assigns a mass of $\frac{2}{5}$ of consumers to the extreme market in which the monopolist is indifferent between charging any price in the support of $x^*$. A share of $\frac{1}{5}$ is assigned to the extreme market where the monopolist is indifferent between prices in $\{1, 2, 3\}$. A mass of $\frac{3}{20}$ is allocated to the extreme market with support $\{1, 3\}$. Finally, the remaining consumers are assigned to the extreme market supported solely on valuation 3.

The key insight from the above segmentation lies in the unique extreme market that includes valuations strictly above the uniform monopoly price, denoted as $x^{\{1,2,3,4\}}$. The mass of consumers assigned to this segment is exactly equal to the uniform lower bound, $\underline{\lambda} = \frac{2}{5}$. This leads to the following necessary condition: a covariate $f$ yields a consumer surplus weakly greater than $u^*$ for all feasible segmentations only if $f_s > \frac{2}{5}$ for every label $s$.

To prove the result, I will show that if $f_s \leq \frac{2}{5}$ for some $s$, then a feasible segmentation exists that yields consumer surplus strictly below $u^*$. Define $\sigma(\cdot \lvert s) = x^{\{1,2,3,4\}}$, and for all $s' \neq s$, define $\sigma(\cdot \lvert s')=\frac{x^*-\sigma(\cdot \lvert s)f_s}{1-f_s}$. This construction satisfies three key properties: (i) Since $f_s \leq \frac{2}{5}$, the segmentation is feasible. (ii) Conditional on observing label $s$, any price in the support of $\sigma(\cdot \lvert s)$ is optimal. Moreover, under the assumption that the monopolist breaks ties against consumers, the chosen price is 4. (iii) The price charged to consumers with any other label $s' \neq s$ is no lower than 3. To see this, suppose instead that the optimal price for some $s'$ is 1 or 2. Then, either of these prices must yield strictly higher profits than 3 for all $s' \neq s$. Furthermore, as $\sigma(\cdot \lvert s)$ is an extreme market, 1 and 2 are an optimal price for that segment. Therefore, 1 or 2 yield strictly higher profits than 3 in the aggregate market $x^*$, which contradicts that 3 is the uniform monopoly price. 

In essence, by allocating all consumers with label $s$ to the extreme market with valuations strictly higher than the uniform monopoly price in the support, I ensure that these consumers receive a price higher than the uniform monopoly price without counterbalancing benefits to consumers with different labels.

To build intuition for why the regulator disallows overly fine covariates that reveal small consumer groups, consider a covariate indicating whether a consumer’s income is below or above a threshold. Let $f = (f_l, f_h)$, with $f_l = f_h = \frac{1}{2}$ denoting equal shares of low- and high-income consumers. Note that this covariate satisfies the uniform lower bound condition, $\underline{\lambda} = \frac{2}{5}$. Moreover, assume that the underlying database indicates that all consumers with valuations greater than or equal to 3 are high-income, and all consumers with valuations strictly below 3 are low-income. In this case, the monopolist will charge a price of 3 to high-income consumers and a price of 1 to low-income consumers. Since low-income consumers receive a price strictly lower than the uniform monopoly price, consumer surplus increases.

Now suppose the monopolist instead has access to a \textit{finer} covariate that reveals both income and gender: $f' = (f'_{ml}, f'_{wl}, f'_{wh}, f'_{mh})$, where the first subscript indicates gender (men or women) and the second income level (low or high). Let $f' = \left(\frac{2}{5}, \frac{1}{10}, \frac{2}{5}, \frac{1}{10}\right)$. This violates the uniform lower bound condition. In the worst-case database consistent with $f'$, each group fully reveals its valuation---1, 2, 3, and 4, respectively---enabling first-degree price discrimination and reducing consumer surplus to zero.
\end{example}

Theorem~\ref{theorem:characterization-WC} yields four immediate corollaries that I describe next. 

\begin{corollary}\label{coro:WC-nonempty}
    $WC\setminus\left\{f^{\emptyset}\right\}$ is non-empty if and only if $\underline{\lambda} < \frac{1}{2}$.
\end{corollary}
Indeed, there is no covariate with finite and full support, other than $f^{\emptyset}$, that assigns a probability strictly greater than $\frac{1}{2}$ to all elements in its support. Therefore, if $\underline{\lambda} > \frac{1}{2}$, $WC\setminus\left\{f^{\emptyset}\right\}$ is empty. Conversely, if $\underline{\lambda} < \frac{1}{2}$, the covariate $f_1=f_2=\frac{1}{2}$ satisfies the uniform lower bound imposed by Theorem~\ref{theorem:characterization-WC}, and thus $f \in WC\setminus\left\{f^{\emptyset}\right\}$.

\begin{corollary}\label{coro:prices}
    $f \in WC$ if and only if for every $\sigma \in \Sigma(f)$, the price charged by the monopolist in every segment $s$, denoted $v_{i^*_s}$, is weakly lower than the uniform monopoly price $v_{i^*}$.
\end{corollary}

If $v_{i^*s} > v_{i^*}$ for every $\sigma \in \Sigma(f)$, then clearly $f \in WC$. The converse follows from the if part of Theorem~\ref{theorem:characterization-WC}, which shows that if a price above the uniform monopoly price maximizes profits in some segment $s$, then $f_{s} \leq \underline{\lambda}$.

To state the next result, define the \textit{join} of two covariates $f_1$ and $f_2$, with label sets $S_1$ and $S_2$, as the covariate $f_1 \vee f_2$, which is a distribution over the product set $S_1 \times S_2$. For example, if gender is represented by $f_1 = (f_m, f_w)$, with $m$ for men and $w$ for women, and income by $f_2 = (f_l, f_h)$, with $l$ for low-income and $h$ for high-income, then $(f_1 \vee f_2)_{s_1s_2}$ denotes the share of consumers with gender $s_1 \in \{m, w\}$ and income $s_2 \in \{l, h\}$. Theorem~\ref{theorem:characterization-WC} immediately implies the following:

\begin{corollary}\label{coro:join}
Suppose $f_1 \notin WC$. Then, for any $f_2 \in F$, we have $f_1 \vee f_2 \notin WC$. Conversely, if $f_1 \vee f_2 \in WC$, then both $f_1 \in WC$ and $f_2 \in WC$.
\end{corollary}

Suppose the $f_1 \notin WC$. Then by Theorem~\ref{theorem:characterization-WC}, $f_{s_1} < \underline{\lambda}$ for some $s_1 \in S_1$. For any covariate $f_2$ with label set $S_2$, it follows that $(f_1 \vee f_2)_{s_1s_2} \leq f_m < \underline{\lambda}$ for all $s_2 \in S_2$. 

Conversely, suppose $f_1 \vee f_2 \in WC$. Then for every $s_1 \in S_1$ and $s_2 \in S_2$, we must have $(f \vee f')_{s_1s_2} > \underline{\lambda}$, which implies $f_{s_1} > \underline{\lambda}$ and similarly $f_{s_2} > \underline{\lambda}$. 

Finally, note that Theorem~\ref{theorem:characterization-WC} can be extended to a setting in which the regulator also seeks policies that are robust to uncertainty about the market distribution. To this end, suppose the regulator only knows that the true market distribution lies within a set $\mathcal{X} \subset \Delta(V)$ and does not assign any prior over this set. For example, if the regulator has observed the uniform monopoly price in the past, then $\mathcal{X}$ may consist of all market distributions for which that price is optimal.

In this case, the regulator evaluates each covariate by its worst-case consumer surplus, computed over all market distributions in $\mathcal{X}$ and their corresponding feasible segmentations. Formally, a policy is worst-case optimal if every admissible covariate belongs to:

\[
WC(\mathcal{X})\equiv\arg \max_{f \in F} \; \inf_{x \in \mathcal{X}} \; \inf_{\sigma \in \Sigma(f)} \; \underline{u}(\sigma,f).
\]

Using Theorem~\ref{theorem:characterization-WC}, we can characterize the set $WC(\mathcal{X})$:

\begin{corollary}\label{coro:robust-market}
A covariate $f$ belongs to $WC(\mathcal{X})$ if and only if $f_s > \sup_{x \in \mathcal{X}} \; \underline{\lambda}(x)$ for every label $s$.
\end{corollary}

The intuition behind Corollary~\ref{coro:robust-market} is straightforward. By Lemma~\ref{lemma:WC=F1}, a covariate belongs to $WC(\mathcal{X})$ if and only if it guarantees a consumer surplus of at least $u^*$ for all $x \in \mathcal{X}$. Moreover, by Theorem~\ref{theorem:characterization-WC}, this condition holds for a given $x$ if and only if $f_s > \underline{\lambda}(x)$ for every label $s$. 

\begin{example}
The set of consumers' valuations is given by $V=\{1,2,3\}$, and the market is $x^*_1=\frac{2}{5}$, $x^*_2=\frac{3}{5}-x^*_3$ and $x^*_3 \in \left[0,\frac{1}{10}\right]$. Figure~\ref{fig:lambda} shows that $\underline{\lambda}$ decreases as $x^*_3$ decreases, and can become arbitrarily small. Note that the uniform monopoly price remains constant at 2 for all values of $x^*_3$. If a covariate leads to consumer surplus below $u^*$, the monopolist must charge a price above 2 in some segment. This requires that the share of consumers with valuation 3 be at least $\frac{1}{3}$. For this to be feasible, it must be possible to pack at least a mass of $\frac{1}{3}$ consumers with valuation 3, which implies $f_s \leq 3x^*_3$. Thus, as $x^*_3$ decreases, so must $f_s$, reducing $\underline{\lambda}$.

\begin{figure}[H]
\centering
\includegraphics[width=0.5\textwidth]{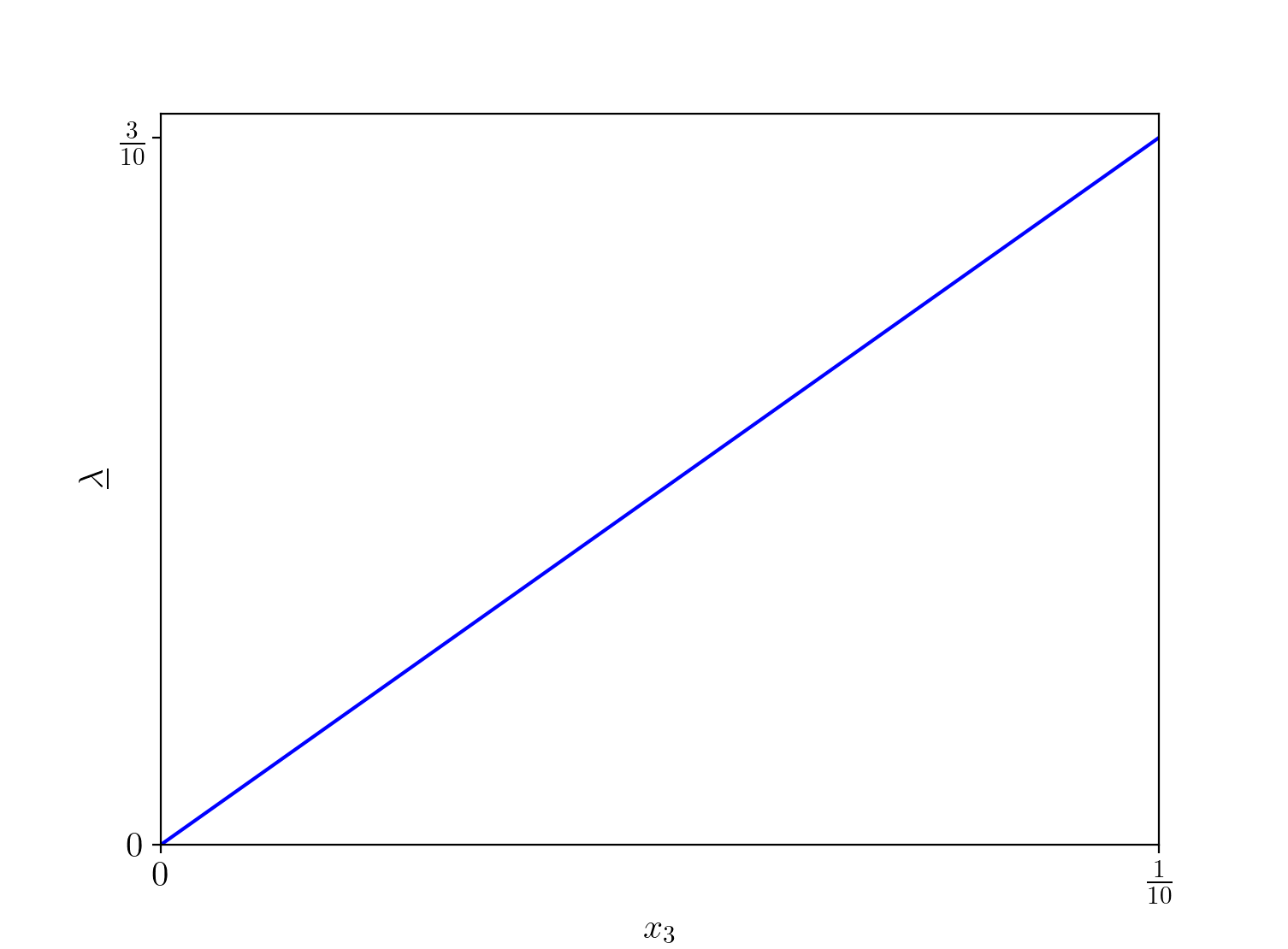}
\caption{Value of $\underline{\lambda}$ for $x_1=\frac{2}{5}$, $x_2=\frac{3}{5}-x_3$ and $x_3 \in \left(0,\frac{1}{10}\right]$}
\label{fig:lambda}
\end{figure}

\end{example}

Figure~\ref{fig:lambda} suggests that $\underline{\lambda}$ decreases as consumers with valuations higher than the uniform monopoly price become less likely and consumers with a valuation equal to the uniform monopoly price become more likely. Proposition~\ref{prop:monotonicity-lambda} formalizes this idea. For this result, and with a slight abuse of notation, let the uniform bound depend on the market, denoted as $\underline{\lambda}(x)$.

\begin{proposition}\label{prop:monotonicity-lambda}
    Let $x,x' \in \Delta(V)$, and suppose that the uniform monopoly price, $v_{i^*}$, is equal in both markets. Furthermore, assume that $\bar{x}_k \geq \bar{x}'_k$ for all $v_k \leq v_{i^*}$, and $\bar{x}'_k \geq \bar{x}_k$ for all $v_k > v_{i^*}$. Then, $\underline{\lambda}(x) \leq \underline{\lambda}(x')$.
\end{proposition}
\begin{proof}
    For any $v_k > v_{i^*}$ and for any $v_j \leq v_{i^*}$ it is true that $v_k(\bar{x}'_k-\bar{x}_k)\geq 0\geq v_j(\bar{x}'_j-\bar{x}_j)$. If I manipulate the later inequality I get that:
    \[
    \frac{v_k\bar{x}'_k}{v_j} + 1 - \bar{x}'_j \geq \frac{v_k\bar{x}_k}{v_j} + 1 - \bar{x}_j,
    \]
    which leads to the desired result.
\end{proof}

\subsection{Dominance Order}\label{sec:dominance}
\noindent After characterizing all worst-case optimal policies available to the regulator, a natural follow-up question arises: among the covariates that meet the uniform lower bound, $\underline{\lambda}$, which ones exhibit a superior performance relative to others? What criteria should be used to compare two covariates that yield the largest worst-case consumer surplus?

To address the questions above, I draw on the insight of \citet{borgers2017no}, who builds on \citet{chung2007foundations}. The latter work analyzes a setting where a revenue-maximizing seller faces non-Bayesian uncertainty about agents' beliefs and identifies conditions under which dominant strategy auctions are worst-case optimal. However, \citet{borgers2017no} shows that there exist other mechanisms that never yield lower expected revenue and sometimes yield higher expected revenue. Consequently, \citet{borgers2017no} proposes a refinement that rules out worst-case optimal mechanisms for which an alternative exists that never yields a lower payoff and sometimes yields a higher payoff.

To illustrate why the regulator should consider the refinement proposed by \citet{borgers2017no}, consider a policy that only allows access to the trivial covariate, i.e., $\mathcal{F}=\left\{f^{\emptyset}\right\}$. $\mathcal{F}$ resembles a policy that prohibits access to consumer information. Furthermore, $\mathcal{F}$ is worst-case optimal. However, suppose that $f \in WC \setminus\left\{f^{\emptyset}\right\}$. By Lemma~\ref{lemma:WC=F1}, it follows that $\underline{u}(\sigma,f) \geq u^*$ for all $\sigma \in \Sigma(f)$. Moreover, if $\underline{u}(\sigma,f) > u^*$ for some $\sigma \in \Sigma(f)$, then there is no reason for the regulator to withhold $f$ from the monopolist.

Building on these observations, I say that $f$ \textit{dominates} $f'$ if $\underline{u}(\sigma,f) \geq \underline{u}(\sigma',f')$ for all $(\sigma,\sigma') \in \Sigma(f) \times \Sigma(f')$, and $\underline{u}(\sigma,f) > \underline{u}(\sigma',f')$ for some $(\sigma,\sigma') \in \Sigma(f) \times \Sigma(f')$. Additionally, I say $f'$ is \textit{dominated} if there exists some $f$ that \textit{dominates} $f'$, and  \textit{undominated} otherwise. 

Let $UD \equiv \left\{f \in WC \; \lvert \; f \; \text{is undominated}\right\}$. The following claim gives a simple expression for $UD$.

\begin{lemma}\label{lemma:UD=F2}
Let $F_2 \equiv \left\{f \in WC \; \lvert \; \underline{u}(\sigma,f) >u^* \; \text{for some} \; \sigma \in \Sigma(f) \right\}$ be the set of covariates that never yield lower consumer surplus than $u^*$ and sometimes higher consumer surplus than $u^*$. Then, $UD \equiv F_2$ if $F_2$ is non-empty, and  $UD \equiv WC$ otherwise.
\end{lemma}

\begin{proof}
    Suppose that $F_2$ is non-empty. Pick any $f \in F_2$ and let $\sigma \in \Sigma(f)$ be the segmentation that yields a consumer surplus strictly higher than $u^*$. For any $f' \in WC$, $\sigma'(\cdot \lvert s)=x^*$ for all $s$ is feasible. Thus, $\underline{u}(\sigma,f) > \underline{u}(\sigma',f')=u^*$; therefore, $f$ is undominated. Conversely, pick any $f \in UD$ and suppose that $f \notin F_2$. Then, $\underline{u}(\sigma,f)=u^*$ for all $\sigma \in \Sigma(f)$; thus, any $f' \in F_2$ dominates $f$, which is a contradiction.

    Now suppose that $F_2$ is empty. Then, for any $f \in WC$, we have that $\underline{u}(\sigma,f)=u^*$ for all $\sigma \in \Sigma(f)$, which implies that any $f$ is undominated, and therefore $UD \equiv W
    C$.
\end{proof}

The second main result of this paper fully characterizes the set $UD$.

\begin{theorem}\label{theorem:characterization-UD}
    Suppose that $f \in WC$. Then, $f$ yields a consumer surplus greater than $u^*$ for some $\sigma \in \Sigma(f)$ if and only if $f_s < \bar{\lambda}$ for some label $s$, where: 
    \[
    \bar{\lambda} =  \max_{v_j < v_{i^*}} \; 1-\frac{v_{i^*}\bar{x}_{i*}-v_j\bar{x}^*_j}{v_{i^*}-v_j}.
    \]
\end{theorem}
\begin{proof}[Only if direction of Theorem~\ref{theorem:characterization-UD}]
     Suppose that $f \in WC$ and $\underline{u}(\sigma,f)>u^*$ for some $\sigma \in \Sigma(f)$. Then, a segment $s$ must exist where the monopolist sets a price lower than $v_{i^*}$. Denote such a price by $v_k$. Since $v_k$ maximizes the monopolist's profits in segment $s$, and the monopolist breaks ties against consumers, it follows that:
    \begin{align}\label{eq:lowbound-sigma-UD}
        v_k\sum_{j=k}^K\sigma(j \lvert s)-v_{i^*}\sum_{j=i^*}^K\sigma(j \lvert s)&=\frac{v_k}{v_{i^*}-v_k}\sum_{j=k}^{i^*-1}\sigma(j \lvert s) + \sum_{j=1}^{i^*-1}\sigma(j \lvert s)-1 > 0. 
    \end{align}
    Furthermore, since $\sigma \in \Sigma(f)$, $\sigma(\cdot \lvert s)f_s \leq x^*$. Then, if I multiply both sides of inequality~\ref{eq:lowbound-sigma-UD} by $f_s$ I get that:
    \begin{align}\label{eq:upbound-f-UD}
        1-\frac{v_{i^*}\bar{x}_{i*}-v_k\bar{x}^*_k}{v_{i^*}-v_k}&=\frac{v_k}{v_{i^*}-v_k}\sum_{j=k}^{i^*-1}x^*_j+\sum_{j=1}^{i^*-1}x^*_j \nonumber \\
        &\geq \left[\frac{v_k}{v_{i^*}-v_k}\sum_{j=k}^{i^*-1}\sigma(j \lvert s)+\sum_{j=1}^{i^*-1}\sigma(j \lvert s)\right]f_s \nonumber \\
        &> f_s.
    \end{align}
    Inequality~\ref{eq:upbound-f-UD} holds for some $v_k < v_{i^*}$, hence: 
    \begin{align*}
        f_s &< \max_{v_k < v_{i^*}} 1-\frac{v_{i^*}\bar{x}_{i*}-v_j\bar{x}^*_k}{v_{i^*}-v_k}= \bar{\lambda}.
    \end{align*}
\end{proof}

I defer the proof of the if part of Theorem~\ref{theorem:characterization-UD} to Appendix~\ref{sec:proofs-UD} and provide a brief sketch here. The goal is to construct a feasible segmentation that yields a consumer surplus greater than $u^*$ for any $f$ that satisfies $f_s<\bar{\lambda}$ for some $s$. Since $f \in WC$, the process is straightforward: in the segment $s$ where $f_s < \bar{\lambda}$, pack as many consumers as needed with valuations lower than $v_{i^*}$. Ensure that the price is less than $v_{i^*}$ while maintaining a positive mass of consumers who enjoy a positive surplus. The condition $f_s < \bar{\lambda}$ ensures the feasibility of such segment. Additionally, by Corollary~\ref{coro:prices}, the prices in the other segments must be lower or equal to $v_{i^*}$. Therefore, consumer surplus must exceed $u^*$.

The following corollary provides sufficient and necessary conditions under which the set $F_2$ non-empty.

\begin{corollary}\label{coro:F2-nonempty}
     $F_2$ is non-empty if and only if $\underline{\lambda}<\frac{1}{2}$ and
    \[
    v_{i^*} \notin \min_{v_j \leq v_{i^*}} \; \frac{\max_{v_i>v_{i^*}} \; v_i \Bar{x}^*_{i}}{v_j}+1-\Bar{x}^*_j.
    \]
\end{corollary}

$F_2$ is non-emppty if and only if $\bar{\lambda}>\underline{\lambda}$. This condition holds true if and only if $v_{i^*} \notin \min_{v_j \leq v_{i^*}} \; \frac{\max_{v_i>v_{i^*}} \; v_i \Bar{x}^*_{i}}{v_j}+1-\Bar{x}^*_j$. 

\begin{corollary} \label{coro:surplus-triangle}
    Suppose that $f \in F_2$, then there is a $\sigma \in \Sigma(f)$ that yields both a consumer surplus and a producer surplus higher than those obtained through uniform pricing.
\end{corollary}

\begin{figure}[H]
    \centering
    \includegraphics[width=0.6\textwidth]{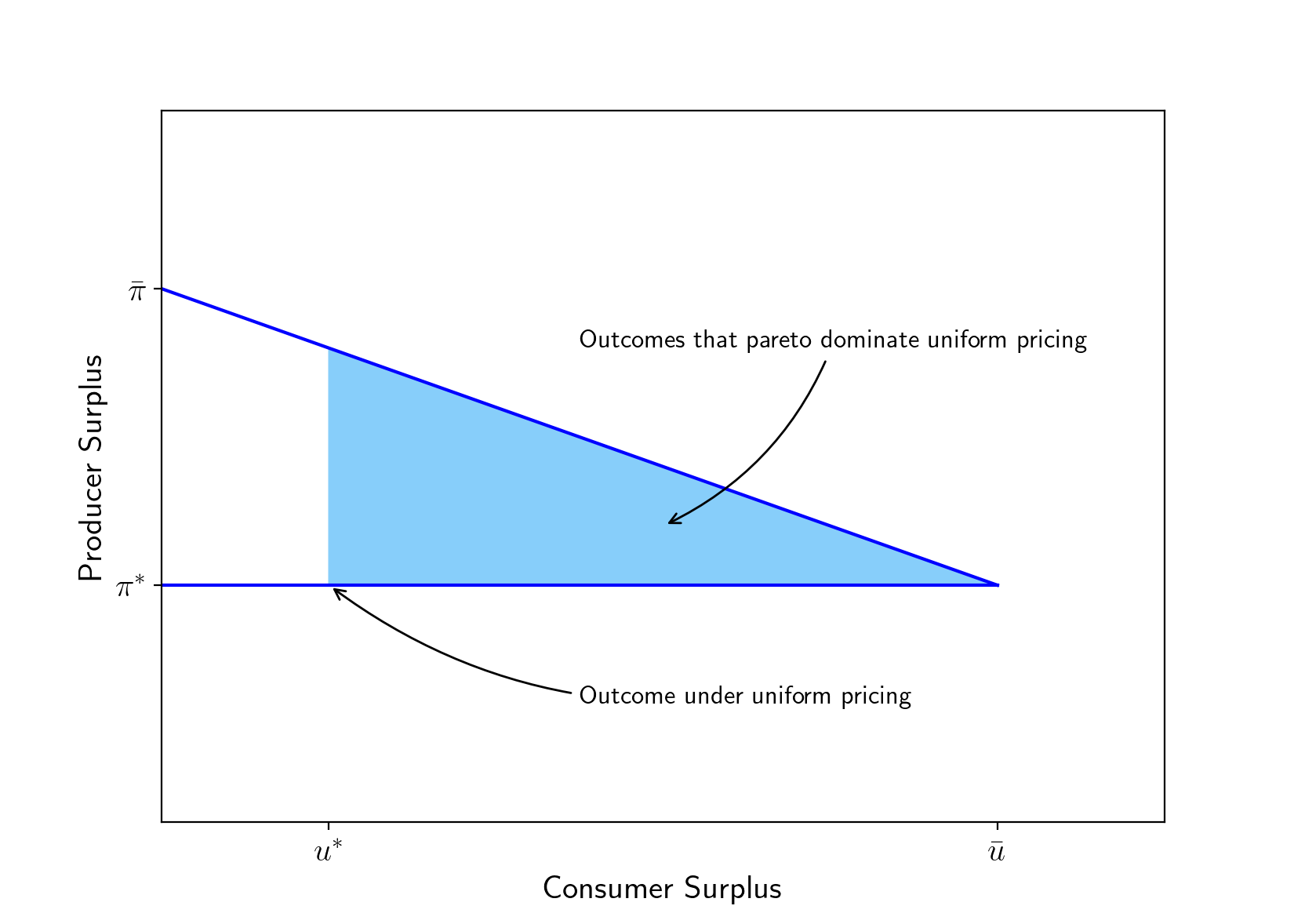}
    \caption{Surplus triangle}
    \label{fig:surplus-triangle}
\end{figure}

Corollary~\ref{coro:surplus-triangle} follows from the proof of the if part of Theorem~\ref{theorem:characterization-UD}. Figure~\ref{fig:surplus-triangle} illustrates Corollary~\ref{coro:surplus-triangle}. The figure shows the surplus triangle derived by \citet{bergemann2015limits}. That is, all combinations of producer and consumer surplus that result from third-degree price discrimination. The point $(\pi^*,u^*)$ corresponds to the producer and consumer surplus achieved with uniform pricing. The shaded area in blue represents the market outcomes, achievable with third-degree price discrimination, that lead to increases in producer and consumer surplus. The covariates characterized by Theorem~\ref{theorem:characterization-UD} always achieve a point inside this blue region. Furthermore, for some feasible segmentation, they achieve the interior of the blue area.

\section{Extensions}\label{sec:extensions}
\noindent In some markets, regulators might prioritize consumers but still care about the potential impact of consumer data on producer surplus. In addressing this concern, I examine the model when the regulator's objective is to maximize a weighted average of consumer and producer surplus, where the weight assigned to consumer surplus is weakly higher than the one assigned to producer surplus. For this purpose, denote by $\pi^*$ the producer surplus achieved with the uniform monopoly price $v_{i^*}$, and by $w^*_{\alpha}$ the $\alpha$-weighted surplus, where $\alpha \in \left[\frac{1}{2},1\right]$ is the weight assigned to $u^*$:
\[
    \pi^* = v_{i^*}\sum_{j=i^*}^{K}x^*_j, \quad  w^*_{\alpha} = \alpha u^* + (1-\alpha) \pi^*.
\]
As a special case, see that when $\alpha=\frac{1}{2}$, the regulator aims to maximize the total surplus, given by $w^*_{\frac{1}{2}}=\frac{1}{2}\sum_{j=i^*}^{K}v_jx^*_j$.

Additionally, let $\underline{\pi}(\sigma^*,f)$ and $\underline{w}_{\alpha}(\sigma^*,f)$ be the producer surplus and the $\alpha$-weighted surplus, respectively, obtained when the monopolist has access to $f$ and he segments the market according to $\sigma^*$:
\[
    \underline{\pi}(\sigma^*,f)=\sum_{s=1}^nf_sv_{i^*_s}\sum_{j=i^*_s}^K\sigma^*(j \lvert s), \quad \underline{w}_{\alpha}(\sigma^*,f)=\alpha\underline{u}(\sigma^*,f)+(1-\alpha)\underline{\pi}(\sigma^*,f),
\]
where $v_{i^*_s}$ is the valuation defined in equation~\ref{eq:monopoly-price}. Then, the set of covariates that achieve the maximum worst-case $\alpha$-weighted surplus is:
\[
WC_{\alpha} \equiv \arg\max_{f \in F} \inf_{\sigma \in \Sigma(f)} \underline{w}_{\alpha}(\sigma,f).
\]

The next result resembles Lemma~\ref{lemma:WC=F1} in the context where the regulator's objective is the $\alpha$-weighted surplus:
\begin{lemma}\label{lemma:WCalpha=F1alpha}
    $f \in WC_{\alpha}$ if and only if  $\underline{w}_{\alpha}(\sigma,f) \geq w^*_{\alpha}$ for all $\sigma \in \Sigma(f)$.
\end{lemma}

I omit the proof of Lemma~\ref{lemma:WCalpha=F1alpha}, as it is essentially the same argument as the one I used to prove Lemma~\ref{lemma:WC=F1}. Then, using Lemma~\ref{lemma:WCalpha=F1alpha}, I can show that any $f$ that attains the maximum worst-case $\alpha$-weighted surplus must satisfy the uniform lower bound defined in Theorem~\ref{theorem:characterization-WC}.
\begin{proposition}\label{prop:characterization-WCalpha}
    Suppose that $f \in WC_{\alpha}$, then $f_s > \underline{\lambda}$ for every label $s$.
\end{proposition}
\begin{proof}
    Towards a contradiction, suppose that $f \in WC_{\alpha}$ but $f_s \leq \underline{\lambda}$ for some label $s$. Then, the segmentation $\sigma$ constructed in the proof of the only if part of Theorem~\ref{theorem:characterization-WC} is feasible given the marginals $f$ and $x^*$. Recall that the price charged by the monopolist in segment $s$ is higher than the uniform monopoly price $v_{i^*}$, and in all other segments, the price is no lower than $v_{i^*}$. Furthermore, segment $s$ contains a positive mass of consumers with a valuation equal to $v_{i^*}$. Hence, $\underline{w}_{\frac{1}{2}}(\sigma,f)<w^*_{\frac{1}{2}}$. Indeed, no consumer with a valuation lower than $v_{i^*}$ gets the good. For all valuations higher or equal to $v_{i^*}$, the mass of consumers that get the good is strictly lower than that of consumers that get the good with the uniform monopoly price.

    If the $\frac{1}{2}$-weighted surplus goes down the following inequality holds:
    \[
    u^*-\underline{u}(\sigma,f^*)>\underline{\pi}(\sigma,f^*)-\pi^*.
    \]
    
    If I multiply the left-hand side of the inequality by $\alpha \in \left[\frac{1}{2},1\right]$ and the right-hand side by $1-\alpha$, the inequality is preserved. Thus, $\underline{w}_{\alpha}(\sigma,f^*)<w^*_{\alpha}$, a contradiction.
\end{proof}

Theorem~\ref{theorem:characterization-WC} and Proposition \ref{prop:characterization-WCalpha} imply the next corollary:
\begin{corollary}\label{coro:characterization-WCalpha}
    The set of covariates that yield the maximum worst-case $\alpha$-weighted surplus is equivalent to the set of covariates that yield the maximum worst-case consumer surplus. 
\end{corollary}
The conclusion follows from initially observing, by Theorem~\ref{theorem:characterization-WC}, that $f \in WC$ if and only if  $f_s>\underline{\lambda}$ for every label $s$. Subsequently, Proposition~\ref{prop:characterization-WCalpha} implies that any $f \in WC_{\alpha}$ also belongs to $WC$. To establish the reverse inclusion, note that if $f \in WC$, it also achieves a consumer surplus weakly greater than $u^*$ for all feasible segmentations. Moreover, under third-degree price discrimination, the producer surplus is always weakly greater than that achieved with uniform pricing. Consequently, $f$ attains an $\alpha$-weighted surplus weakly greater than $w^*_{\alpha}$ for all feasible segmentations, and thus $f \in WC_{\alpha}$. 

Finally, I will show that the set of undominated covariates that achieve the maximum worst-case $\alpha$-weighted surplus also remains the same. Let:
\[
UD_{\alpha} \equiv \left\{f \in WC_{\alpha} \; \lvert \; f \; \text{is undominated}\right\},
\]
where $f$ is undominated if there is no other $f' \in WC_{\alpha}$ that never yields lower $\alpha$-weighted surplus than $f$ and sometimes higher $\alpha$-weighted surplus than $f$. 

Once more, the next result resembles Lemma~\ref{lemma:UD=F2} in the context where the regulator's objective is the $\alpha$-weighted surplus:
\begin{lemma}\label{lemma:UDalpha=F2alpha}
    Let $F_{\alpha} \equiv \left\{f \in WC_{\alpha} \; \lvert \; \underline{w}_{\alpha}(\sigma,f) > w^*_{\alpha} \; \text{for some} \; \sigma \in \Sigma(f) \right\}$ be the set of covariates that never yield lower $\alpha$-weighted surplus than $w^*_{\alpha}$ and sometimes higher $\alpha$-weighted surplus than $w^*_{\alpha}$. Then, $UD_{\alpha} \equiv F_{\alpha}$ if $F_{\alpha}$ is non-empty, and  $UD_{\alpha} \equiv WC_{\alpha}$ otherwise.
\end{lemma}
I exclude the proof of Lemma~\ref{lemma:UDalpha=F2alpha} as it is the same argument as the one I used to prove Lemma~\ref{lemma:UD=F2}. Furthermore, it is still true that any $f$ that never yields lower $\alpha$-weighted surplus than $w^*_{\alpha}$ and sometimes higher $\alpha$-weighted surplus than $w^*_{\alpha}$ must attach a mass lower than $\bar{\lambda}$ to some label:
\begin{proposition}\label{prop:characterization-UDalpha}
    Suppose that $f \in F_{\alpha}$, then $f^*_s < \bar{\lambda}$ for some label $s$.
\end{proposition}
\begin{proof}
    Suppose that $f \in WC_{\alpha}$ and  $\underline{w}_{\alpha}(\sigma,f) > w^*_{\alpha}$ for some $\sigma \in \Sigma(f)$. For this inequality to hold, there must be a segment where the price the monopolist charges differs from $v_{i^*}$. Additionally, by Corollary~\ref{coro:prices}, in every segment, the price charged by the monopolist is weakly lower than $v_{i^*}$. Then, for some segment $s$, the price must be strictly lower than $v_{i^*}$. Moreover, if I follow the argument used to show the only if part of Theorem~\ref{theorem:characterization-UD} I will get that whenever a price lower than $v_{i^*}$ is an optimal price for segment $s$, then $f_s<\bar{\lambda}$.
\end{proof}

Theorem~\ref{theorem:characterization-UD} and Proposition~\ref{prop:characterization-UDalpha} yield the next corollary
\begin{corollary}\label{coro:characterization-UDalpha}
    The set of covariates that never yield lower $\alpha$-weighted surplus than $w^*_{\alpha}$ and sometimes higher $\alpha$-weighted surplus than $w^*_{\alpha}$ is equivalent to the set of covariates that never yield lower consumer surplus than $u^*$ and sometimes higher consumer surplus than $u^*$. 
\end{corollary}
The assertion that any covariate that belongs to $F_{\alpha}$ is also part of the set $UD$ can be derived from Theorem~\ref{theorem:characterization-UD} and Proposition~\ref{prop:characterization-UDalpha}. The reverse inclusion is established through a similar argument as delineated in Corollary \ref{coro:characterization-WCalpha}: If $f$ never achieves a lower producer surplus than the one attained with uniform pricing, and, additionally, $f$ achieves a higher consumer surplus than $u^*$ for some segmentation. It also attains a higher $\alpha$-weighted surplus for that specific segmentation than $w^*_{\alpha}$.

\section{Conclusion}\label{sec:conclusions}
\noindent In this paper, I took a normative approach to studying the regulation of data markets. I analyzed a setting where a regulator determines what data a firm can access to maximize consumer surplus. Furthermore, I assumed this regulator had limited knowledge of the relevant environment. Notably, she possesses non-Bayesian uncertainty regarding the correlation between data and consumers' willingness to pay.

Through this paper, I assumed that the monopolist uses consumer data for price discrimination. However, a promising direction for future work is to explore how the policies the regulator implements change when a firm uses the data for multiple purposes besides price discrimination. For instance, firms may use consumer data to offer consumers personalized products that better match their preferences.In this setting, it is still reasonable to assume that regulators have limited knowledge of the products a firm will recommend and the prices consumers will face. In particular, this will highly depend on the correlation between the data and willingness to pay. Therefore, the modeling ideas used in this paper can be helpful in such a setting: The regulator evaluates a covariate across all feasible joint distributions of consumers' characteristics and valuations for each product.

\begin{appendix}
\section{Proof of the Only If Direction of Theorem~\ref{theorem:characterization-WC}}\label{sec:proofs-WC}
\subsection{Greedy Algorithm to Construct Extreme Segmenations}
\noindent The proof of the only if part of Theorem~\ref{theorem:characterization-WC} builds on the algorithm proposed by \citet{bergemann2015limits} that segments $x^*$  by means of a ``greedy'' procedure and creates segments which are extreme markets. 

Before formally presenting the algorithm, I define an \textit{extreme market}. Let $\mathcal{V}$ represent the set of non-empty subsets of $V=\{v_1,...,v_K\}$. For every subset $S \in \mathcal{V}$, \citet{bergemann2015limits} refer to $x^{S} \in \Delta(V)$ as an \textit{extreme market} if it satisfies the properties that:
\begin{enumerate}[label=(\roman*)]
    \item no consumer has a valuation outside the set $S$, and
    \item the monopolist is indifferent between charging any price inside the set $S$.
\end{enumerate}

\citet{bergemann2015limits} show that with the above conditions, it is easy to find an analytical expression for $x^S$. Writing $\min S$ and $\max S$ for the smallest and greatest element of $S$, respectively, and denoting by $\mu(v_i,S)$ the smallest element of $S$ which is strictly greater than $v_i$, we must have:
\begin{equation}\label{eq:extreme-markets}
    x_{i}^{S}= \begin{cases}
    0 \;  &\text{if} \; v_i \notin S;\\
    \min S\left(\frac{1}{v_i}-\frac{1}{\mu(v_i,S)}\right) \; &\text{if} \; v_i \in S \; \text{and} \; v_i \neq \max S; \\
    \frac{\min S}{\max S} \; &\text{if} \; v_i=\max S.
    \end{cases}
\end{equation}

Having defined $x^S$, I can now describe the steps of the greedy algorithm exactly as outlined by \citet{bergemann2015limits}: At the end of the iteration $l \geq 0$, the ``residual'' market of valuations not yet assigned to a segment is $x^{l}$, with $x^0=x^*$, and the support of this residual is defined to be $S_l=\supp x^l$. I now describe what happens at iteration $l$, taking as inputs $\left(x^{l-1},S_{l-1}\right)$. If $x^{l-1}=x^{S_{l-1}}$, then let $\alpha^{l}=1$ and $x^{l}=0$. Otherwise, define the projection $z(t)$ as:
\begin{equation}\label{eq:projection}
    z(t) = x^{S_{l-1}}+t \cdot \left(x^{l-1}-x^{S_{l-1}}\right).
\end{equation}

For this projection, let $\hat{t}=\inf\{t \geq 0 \lvert z_i(t) < 0 \; \text{for some $i$}\}$, such that $z_{i}\left(\hat{t}\right) \geq 0$ for all $i$ and $z_i\left(\hat{t}\right)=0$ for some $i$. See that $\hat{t}$ is well defined as there is some $i$ for which $z_i(t)$ eventually hits zero. That is, there is some $i$ for which $x_{i}^{l-1} < x_{i}^{S_{l-1}}$. Indeed, since $x^{l-1} \neq x^{S_{l-1}}$, having $x_i^{l-1} \geq x_{i}^{S_{l-1}}$ for all $i$ would violate probabilities in $x^{l-1}$ summing to one. Furthermore, as all elements of $x^{l-1}$, and all elements of $x^{S_{l-1}}$, add up to one, it follows that $\sum_{j=1}^Kz_j\left(\hat{t}\right)=1$. Therefore, $z\left(\hat{t}\right) \in \Delta(V)$. The next residual market $x^l$ is defined to be $z\left(\hat{t}\right)$, and $\alpha^l$ is defined by:
\[
x^{l-1} = \alpha^{l} \cdot x^{S_{l-1}} + (1-\alpha^{l}) \cdot x^{l}.
\]

Now inductively define $S_l=\supp x^l$, which is a strict subset of $S_{l-1}$. The inductive hypothesis is that:
\begin{equation}\label{eq:segmentation}
x^*=\sum_{j=1}^{l}\alpha^j\prod_{i=1}^{j-1}(1-\alpha^i)\cdot x^{S_{j-1}}+\prod_{i=1}^{l}(1-\alpha^i)\cdot x^l,
\end{equation}
meaning that the aggregate market is segmented by the $x^{S_j}$ extreme markets for $j<l$ and the residual market. This property is trivially satisfied for the base case $l=1$ (with the convention that the empty product is equal to 1), and the choice of $\alpha^l$ guarantees that if the inductive hypothesis holds at $l-1$, it will continue to hold at $l$ as well. The algorithm terminates at iteration $L+1$ when $x^{L}=x^{S_L}$, which certainly has to be the case when $\lvert S_L \lvert=1$. Therefore, $x^*$ is segmented by the extreme markets $\{x^{S_l}\}_{l=0}^{L}$ and the mass of consumers assigned to each segment is $y\left(x^{S_l}\right)=\alpha^{l+1}\prod_{j=1}^{l}(1-\alpha^j)$.

\subsection{Notation}
\noindent Let $v_{\bar{i}}=\max \left\{\arg \max_{v_j > v_{i^*}} \; v_j\bar{x}^*_j\right\}$ and $v_{\underline{i}}=\min \left\{\arg \min_{v_j \leq v_{i^*}} \frac{v_{\bar{i}}\bar{x}^*_{\bar{i}}}{v_j}+1-\bar{x}^*_j\right\}$.
Therefore, $\underline{\lambda}= \frac{v_{\bar{i}}\bar{x}^*_{\bar{i}}}{v_{\bar{i}}}+1-\bar{x}^*_{\bar{i}}$.

\subsection{Outline of the Proof}
\begin{itemize}
    \item \textbf{Sep 1:} In the first step, I show that if $x^*$ is segmented using the greedy algorithm, then, if at some iteration there is a positive mass to assign of consumers with valuations higher than $v_{\bar{i}}$, there must also be a positive mass to assign of consumers with valuation $v_{\bar{i}}$. That is, if $x^l_i>0$ for some $v_i > v_{\bar{i}}$, then $x^l_{\bar{i}}>0$. Similarly, if there is a positive mass to assign of consumers with valuation $v_{\bar{i}}$, then there is a positive mass to assign of consumers with valuation $v_{\underline{i}}$; i.e., if $x^l_{\bar{i}}>0$, then $x^l_{\underline{i}}>0$ . Finally, at some point, the algorithm will run out of mass of consumers with valuations lower than $v_{\underline{i}}$, but there will be a positive mass of consumers with valuation $v_{\bar{i}}$; i.e., if $x^{l-1}_i>0$ for some $v_i < v_{\underline{i}}$, then $x^l_{\bar{i}}>0$.
    \item \textbf{Step 2:} Then, I use \textbf{Step 1} to show that the algorithm assigns a total mass of consumers equal to $\underline{\lambda}$ to all extreme markets that contain $v_{\bar{i}}$ in its support.
    \item \textbf{Step 3:} Finally, I use \textbf{Step 2} to show that if $f_s \leq \underline{\lambda}$ I can construct a feasible segmentation that leads to a consumer surplus strictly lower than $u^*$.
\end{itemize}

\subsection{Step 1}
\begin{lemma}\label{lemma:ibar-fillslast}
    Suppose that $x^l_i>0$ for some $v_i > v_{\bar{i}}$. Then, $x^l_{\bar{i}}>0$.
\end{lemma}

\begin{proof}[Proof of Lemma~\ref{lemma:ibar-fillslast}]
The inductive hypothesis is that $x_{\bar{i}}^{l-1}>0$ and $v_{\bar{i}}\bar{x}_{\bar{i}}^{l-1} > v_j\bar{x}_j^{l-1}$ for all $v_j \in S_{l-1}\setminus\{v_1,v_2,...,v_{\bar{i}}\}$. I will use this inductive hypothesis to show that if $x^l_i>0$ for some $v_i > v_{\bar{i}}$, then $x_{\bar{i}}^{l}>0$ and $v_{\bar{i}}\bar{x}_{\bar{i}}^{l} > v_j\bar{x}_j^{l}$ for all $v_j \in S_{l}\setminus\{v_1,v_2,...,v_{\bar{i}}\}$.

Since $x^l \neq 0$, $x^{l}$ is defined by the projection $z(t)$ (equation~\ref{eq:projection}) evaluated at $\hat{t}>0$. Therefore, for any $v_j \in S_{l-1}$, $v_j\bar{x}^{l}_j$ is given by:
\begin{align}
    v_j\bar{x}^l_{j}&=\hat{t}v_j\bar{x}_{j}^{l-1}-\left(\hat{t}-1\right)v_j\bar{x}_{j}^{S_{l-1}} \nonumber \\
    &=\hat{t}v_j\bar{x}_{j}^{l-1}-\left(\hat{t}-1\right)\min S_{l-1},
    \label{eq:ibar-projection}
\end{align}
where for the second equality, I used the analytical expression for extreme markets (equation~\ref{eq:extreme-markets}). Note that, by the inductive hypothesis, $\hat{t}v_{\bar{i}}\bar{x}_{\bar{i}}^{l-1}>\hat{t}v_j\bar{x}_{j}^{l-1}$ for all $v_j \in S_{l-1}\setminus\{v_1,v_2,...,v_{\bar{i}}\}$. Additionally, the second term of equation~\ref{eq:ibar-projection} is the same for all valuations in $S_{l-1}$. Therefore, $v_{\bar{i}}\bar{x}_{\bar{i}}^{l} > v_j\bar{x}_j^{l}$ for all $v_j \in S_{l-1}\setminus\{v_1,v_2,...,v_{\bar{i}}\}$, and, as $S_l \subset S_{l-1}$, the inequality is also true for all for all $v_j \in S_{l}\setminus\{v_1,v_2,...,v_{\bar{i}}\}$. In particular, $v_{\bar{i}}\bar{x}^{l}_{\bar{i}}>v_k\bar{x}^{l}_{k}$, where $v_k$ is the smallest valuation strictly higher than $v_{\bar{i}}$ in $S_l$. See that $v_k$ exists as I assumed $x^l_i>0$ for some $v_i > v_{\bar{i}}$. Moreover, since $v_k>v_{\bar{i}}$, it follows that $v_k\bar{x}^{l}_{\bar{i}}>v_k\bar{x}^{l}_{k}$, which leads to $\bar{x}^{l}_{\bar{i}}-\bar{x}^{l}_k=x^l_{\bar{i}}>0$.

Finally, the base case satisfies that $x^{0}_{\bar{i}}>0$ and $v_{\bar{i}}\bar{x}_{\bar{i}}^{0} > v_j\bar{x}_j^{0}$ for all $v_j \in S_{0}\setminus\{v_1,v_2,...,v_{\bar{i}}\}$, where $x^0=x^*$. Thus, I can conclude that if $x^l_i>0$ for some $v_i>v_{\bar{i}}$, then $x^l_{\bar{i}}>0$.
\end{proof}

\begin{lemma}\label{lemma:ibar-before-iunderline}
    Suppose that $x^l_{\bar{i}}>0$. Then, $x^l_{\underline{i}}>0$.
\end{lemma}

\begin{proof}[Proof of Lemma~\ref{lemma:ibar-before-iunderline}]
The inductive hypothesis is that $x^{l-1}_{\underline{i}}>0$ and: 
\[
v_{\underline{i}}=\min \left\{\arg \min_{v_j \in S_{l-1}}\;\frac{v_{\bar{i}}\bar{x}^{l-1}_{\bar{i}}}{v_j}+1-\bar{x}^{l-1}_j\right\}.
\]

I will show that if $x_{\bar{i}}^l>0$, then $x_{\underline{i}}^l>0$ and $v_{\underline{i}}=\min \left\{\arg \min_{v_j \in S_l}\;\frac{v_{\bar{i}}\bar{x}^{l}_{\bar{i}}}{v_j}+1-\bar{x}^{l}_j\right\}$.

If I use the analytical expression for extreme markets (equation~\ref{eq:extreme-markets}), it is easy to show that for all $v_j \in S_{l-1}$:
\begin{align*}
    \frac{v_{\bar{i}}\bar{x}^{S_{l-1}}_{\bar{i}}}{{v_j}} + 1-\bar{x}^{S_{l-1}}_{j}&=\frac{v_{\bar{i}}}{v_j}\frac{\min S_{l-1}}{v_{\bar{i}}} + 1-\frac{\min S_{l-1}}{v_j}=1.\\
\end{align*}

Furthermore, since $x^{l} \neq 0$, $x^{l}$ is defined by the projection $z(t)$ (equation~\ref{eq:projection}) evaluated at $\hat{t}>0$: 
\begin{align}\label{eq:ibar-iunderline-projection}
     \frac{v_{\bar{i}}x^{l}_{\bar{i}}}{v_j} + 1-\bar{x}^{l}_{j} = \hat{t} \left(\frac{v_{\bar{i}}x^{l-1}_{\bar{i}}}{v_j} + 1-\bar{x}^{l-1}_{j}\right) - (\hat{t}-1)\left(\frac{v_{\bar{i}}x^{S_{l-1}}_{\bar{i}}}{v_j} + 1-\bar{x}^{S_{l-1}}_{j}\right).
\end{align}

Note that, by the inductive hypothesis, $v_{\underline{i}}$ is the smallest element in $S_{l-1}$ to minimize the first term in parentheses, and before I showed that the second term in parenthesis is equal to 1 for all $v_j \in S_{l-1}$. Then, $v_{\underline{i}}$ is the smallest element in $S_{l-1}$ that minimizes~\ref{eq:ibar-iunderline-projection}. Furthermore, as $S_l \subset S_{l-1}$, the value of~\ref{eq:ibar-iunderline-projection} is strictly lower under $v_{\underline{i}}$ compared to any lower valuation in $S_l$, and weakly lower compared to any higher valuation in $S_l$. In particular, denote by $v_k$ the smallest element in $S_{l}$ which is strictly greater than $v_{\underline{i}}$, where $v_k$ exists as $x_{\bar{i}}^l>0$ and $v_{\bar{i}}>v_{\underline{i}}$. Then, see that:
\begin{align}
    \frac{v_{\bar{i}}\bar{x}^{l}_{\bar{i}}}{v_k} + 1-\bar{x}^{l}_{k} &\geq \frac{v_{\bar{i}}\bar{x}^{l}_{\bar{i}}}{v_{\underline{i}}} + 1-\bar{x}^{l}_{\underline{i}} \nonumber \\
    \bar{x}^{l}_{\underline{i}}-
   x_{\underline{i}}^{l}&\geq v_{\bar{i}}\bar{x}^{l}_{\bar{i}}\left(\frac{1}{v_{\underline{i}}}-\frac{1}{v_{k}}\right).\label{eq:ibar-iunderline-inequality}
\end{align}

Since $x^l_{\bar{i}}>0$, the left-hand side of~\ref{eq:ibar-iunderline-inequality} is strictly positive, therefore, $x_{\underline{i}}^l>0$.

Finally, the base case satisfies that $v_{\underline{i}}=\min\left\{\arg\min_{v_j \leq v_{i^*}} \; \frac{v_{\bar{i}}x^{0}_{\bar{i}}}{v_j} + 1-\bar{x}^{0}_{j}\right\}$ and $x_{\underline{i}}^0>0$, where $x^{0}=x^*$. The only issue is that $v_{\bar{i}}$ is the smallest valuation in the $\arg \min$ when the $\min$ is taken across all valuations lower or equal than $v_{i^*}$, and I need the $\min$ to be across all valuations in $S_0 = \supp x^*$. However, $v_{\bar{i}}$ is still the smallest valuation in the $\arg \min$ when the $\min$ is taken across $S_0$. Indeed, towards a contradiction, suppose that $v_j > v_{i^*}$ and: 
\begin{align*}
    \frac{v_{\bar{i}}x^*_{\bar{i}}}{v_{j}}+1-\bar{x}^{*}_{j} &\leq \frac{v_{\bar{i}}x^*_{\bar{i}}}{v_{i^*}}+1-\bar{x}^*_{i^*} \\
    v_{i^*}v_{j}\bar{x}^*_{i^*} &\leq (v_j-v_{i^*})v_{\bar{i}}x^*_{\bar{i}} + v_{i^*}v_j\bar{x}^*_j \\
    v_{i^*}v_{j}\bar{x}^*_{i^*} &< (v_j-v_{i^*})v_{i^*}\bar{x}^*_{i^*}+v_{i^*}v_{i^*}\bar{x}^*_{i^*}\\
    \bar{x}^*_{i^*} &< \bar{x}^*_{i^*},
\end{align*}
where the strict inequality follows from the fact that $v_{i^*}$ is the highest optimal price; therefore, $v_{i^*}\bar{x}^*_{i^*} > v_j\bar{x}^*_j$. Then, $\frac{v_{\bar{i}}x^*_{\bar{i}}}{v_{j}}+1-\bar{x}^{*}_{j} >\frac{v_{\bar{i}}x^*_{\bar{i}}}{v_{i^*}}+1-\bar{x}^{*}_{i^*} \geq \frac{v_{\bar{i}}x^*_{\bar{i}}}{v_{\underline{i}}}+1-\bar{x}^{*}_{\underline{i}}$.
\end{proof}

\begin{lemma}\label{lemma:iunderline-fillslast}
    Suppose that $x^{l-1}_i>0$ for some $v_i < v_{\underline{i}}$. Then, $x^l_{\bar{i}}>0$.
\end{lemma}

\begin{proof}[Proof of Lemma~\ref{lemma:iunderline-fillslast}]
The inductive hypothesis is that $x_{\bar{i}}^{l-1}>0$. I will show that if $x^{l-1}_i>0$ for some $v_i < v_{\underline{i}}$, then $x_{\bar{i}}^{l}>0$.

Since $x_{\bar{i}}^{l-1}>0$, Lemma~\ref{lemma:ibar-before-iunderline} implies that $x_{\underline{i}}^{l-1}>0$ and: 
\begin{equation}\label{eq:viunderline}
v_{\underline{i}}=\min\left\{\arg \min_{v_j \in S_{l-1}} \; \frac{v_{\bar{i}}\bar{x}^{l-1}_{\bar{i}}}{v_j} + 1-\bar{x}^{l-1}_{j}\right\}.
\end{equation}

Equation~\ref{eq:viunderline} together with the assumption that  $x^{l-1}_i>0$ for some $v_i < v_{\underline{i}}$ implies that $x^{l-1} \neq x^{S_{l-1}}$. Otherwise,  there is a valuation $v_i < v_{\underline{i}}$ and $v_i \in \arg\min_{v_j \in S_{l-1}} \frac{v_{\bar{i}}\bar{x}^{S_{l-1}}_{\bar{i}}}{v_j} + 1-\bar{x}^{S_{l-1}}_{j}$. Indeed, in the proof of Lemma~\ref{lemma:ibar-before-iunderline} I showed the term inside the $\arg\min$ equals 1 for all $v_j \in S_{l-1}$. The fact that $x^{l-1} \neq x^{S_{l-1}}$ implies that $x^{l} \neq 0$, and therefore $x^l$ is defined by the projection $z(t)$ (equation~\ref{eq:projection}) evaluated at $\hat{t}>0$.

Denote by $v_k$ the greatest element in $S_{l-1}$ that it is strictly smaller than $v_{\underline{i}}$, where $v_k$ exists as I assumed that $x^{l-1}_i>0$ for some $v_i < v_{\underline{i}}$. Equation~\ref{eq:viunderline} leads to:
\begin{align}
    \frac{v_{\bar{i}}\bar{x}^{l-1}_{\bar{i}}}{v_k} + 1-\bar{x}^{l-1}_{k} &> \frac{v_{\bar{i}}\bar{x}^{l-1}_{\bar{i}}}{v_{\underline{i}}} + 1-\bar{x}^{l-1}_{\underline{i}} \nonumber \\
    v_{\bar{i}}\bar{x}^{l-1}_{\bar{i}} &> \left(\frac{1}{v_{k}}-\frac{1}{v_{\underline{i}}}\right)^{-1}x_{k}^{l-1}.\label{eq:iunderline-inequality}
\end{align}

Furthermore, the fact that $x^{l}=z\left(\hat{t}\right)$ implies that:
\begin{align*}
    x_{k}^{l}&=\hat{t}x_{k}^{l-1}-\left(\hat{t}-1\right)x_{k}^{S_{l-1}} \\
    &=\hat{t}x_{k}^{l-1}-\left(\hat{t}-1\right)\min S_{l-1}\left(\frac{1}{v_{k}}-\frac{1}{v_{\underline{i}}}\right), \quad \text{and} \\
    \bar{x}_{\bar{i}}^{l}&=\hat{t}\bar{x}_{\bar{i}}^{l-1}-\left(\hat{t}-1\right)\bar{x}_{\bar{i}}^{S_{l-1}} \\
    &= \hat{t}\bar{x}_{\bar{i}}^{l-1}-\left(\hat{t}-1\right)\frac{\min S_{l-1}}{v_{\bar{i}}}.
\end{align*}
Since $x_{k}^{l} \geq 0$, $\hat{t}\left(\frac{1}{v_{k}}-\frac{1}{v_{\underline{i}}}\right)^{-1}x_{k}^{l-1} \geq \left(\hat{t}-1\right)\min S_{l-1}$, then inequality~\ref{eq:iunderline-inequality} implies that $\hat{t}v_{\bar{i}}\bar{x}_{\bar{i}}^{l-1}>\left(\hat{t}-1\right)\min S_{l-1}$, which leads to $\bar{x}_{\bar{i}}^{l}>0$. Moreover, if $\bar{x}_{\bar{i}}^{l}>0$, then $x^l_{i}>0$ for some $v_i \geq v_{\bar{i}}$. If $v_i=v_{\bar{i}}$ I get the desired result. If $v_i > v_{\bar{i}}$, Lemma~\ref{lemma:ibar-fillslast} implies that also $x_{\bar{i}}^{l}>0$, which also gives me the desired result.

Finally, for the base case, it is true that $x^0_{\bar{i}}>0$. Therefore, I can conclude that if $x^{l-1}_i>0$ for some $v_i < v_{\underline{i}}$, then $x_{\bar{i}}^{l}>0$. 
\end{proof}

\subsection{Step 2}
\begin{lemma}\label{lemma:weight=lambda}
    Let $T>0$ be the first step at which $x_{\bar{i}}^{T}=0$. Then, $\sum_{j=0}^{T-1}y\left(x^{S_j}\right)=\underline{\lambda}$.
\end{lemma}

\begin{proof}[Proof of Lemma~\ref{lemma:weight=lambda}]
    Denote by $\{l_{1},l_{2},...,l_{N}\}$ the sub-sequence of steps lower than $T$ such that at step $l_n$, $\min S_{l_n}=\min \supp x^{l_n}$ becomes for the first time the minimum valuation. That is, $\min S_{l_n} \neq \min S_l$ for all $l < l_n$. Since $S_0=\supp x^*$ and $S_l \subset S_{l-1}$, is immediate that $l_1=0$, $\min S_{l_1}=\min \supp x^*$ and $\min S_{l_1} < \min S_{l_2} <\dots<\min S_{l_N}$. Additionally, by assumption, $x_{\bar{i}}^l>0$ for all $l<T$, by Lemma~\ref{lemma:ibar-before-iunderline}, $x^l_{\underline{i}}>0$ for all $l<T$, and,  by Lemma~\ref{lemma:iunderline-fillslast}, if $v_i < v_{\underline{i}}$, then $x^l_{i}=0$ for some $l< T$. Therefore, $\min S_{l_N}=v_{\underline{i}}$ and for all $j \in \{l_{n},...,l_{n+1}-1\}$:
    \begin{align}
        1-\bar{x}_{\underline{i}}^{S_j}&=\sum_{k=1}^{\underline{i}-1}x_k^{S_j}=\min S_{l_n}\left(\frac{1}{\min S_{l_n}}-\frac{1}{v_{\underline{i}}}\right) \nonumber \\
        &=1-\frac{\min S_{l_n}}{v_{\underline{i}}}, \quad \text{and} \label{eq:xbar-iunderline}\\
        \bar{x}_{\bar{i}}^{S_j}&=\sum_{k=\bar{i}}^Kx_k^{S_j}=\frac{\min S_{l_n}}{v_{\bar{i}}}, \label{eq:xbar-ibar}
    \end{align}
    where I am using the analytical expression for extreme markets (equation~\ref{eq:extreme-markets}).

    Next, see that if I evaluate equation~\ref{eq:segmentation} at step $T$ and I add it across all valuations strictly lower than $v_{\underline{i}}$ I get that:
    \begin{align*}
        1-\bar{x}^*_{\underline{i}}=\sum_{k=1}^{\underline{i}-1}x^*_k&=\sum_{j=1}^{T}\alpha^j\prod_{i=1}^{j-1}(1-\alpha^i)\sum_{k=1}^{\underline{i}-1}x_k^{S_{j-1}}+\prod_{i=1}^{T}(1-\alpha^i)\sum_{k=1}^{\underline{i}-1}x_k^{l}\\
        &=\sum_{n=1}^{N}\sum_{j=l_n+1}^{l_{n+1}}\alpha^j\prod_{i=1}^{j-1}(1-\alpha^i)\sum_{k=1}^{\underline{i}-1}x_k^{S_{j-1}}\\
        &=\sum_{n=1}^{N}\sum_{j=l_n+1}^{l_{n+1}}\alpha^j\prod_{i=1}^{j-1}(1-\alpha^i)\left(1-\frac{\min S_{l_n}}{v_{\underline{i}}}\right) \\
        &=\sum_{n=1}^{N}\sum_{j=l_n+1}^{l_{n+1}}\alpha^j\prod_{i=1}^{j-1}(1-\alpha^i)-\sum_{n=1}^{N}\frac{\min S_{l_n}}{v_{\underline{i}}}\sum_{j=l_n+1}^{l_{n+1}}\alpha^j\prod_{i=1}^{j-1}(1-\alpha^i),
    \end{align*}
    where I am using the convention that $l_{N+1}=T$, in the second equality I used the assumption that $x_{\bar{i}}^T=0$, hence Lemma~\ref{lemma:iunderline-fillslast} implies that $x_{k}^T=0$ for all $v_k < v_{\underline{i}}$, and in the third equality I used equation~\ref{eq:xbar-iunderline}. Furthermore, recall that the mass of consumers assigned to the extreme market $x^{S_j}$ is equal to $y\left(x^{S_j}\right)=\alpha^{j+1}\prod_{i=1}^j(1-\alpha^i)$. Thus:
    \begin{equation}
        \sum_{j=0}^{T-1}y\left(x^{S_j}\right)=1-\bar{x}^*_{\underline{i}}+\sum_{n=1}^{N}\frac{\min S_{l_n}}{v_{\underline{i}}}\sum_{j=l_n+1}^{l_{n+1}}\alpha^j\prod_{i=1}^{j-1}(1-\alpha^i). \label{eq:weight}
    \end{equation}

    If I now evaluate equation~\ref{eq:segmentation} at step $T$ and I add it across all valuations greater or equal than $v_{\bar{i}}$ I get that:
    \begin{align}
        \bar{x}^*_{\bar{i}}=\sum_{k=\bar{i}}^{K}x^*_k&=\sum_{n=1}^{N}\sum_{j=l_n+1}^{l_{n+1}}\alpha^j\prod_{i=1}^{j-1}(1-\alpha^i)\sum_{k=\bar{i}}^{K}x_{k}^{S_{j-1}}+\prod_{i=1}^{T}(1-\alpha^i)\sum_{k=\bar{i}}^{K}x_{k}^{l} \nonumber \\
        &=\sum_{n=1}^{N}\frac{\min S_{l_n}}{v_{\bar{i}}}\sum_{j=l_n+1}^{l_{n+1}}\alpha^j\prod_{i=1}^{j-1}(1-\alpha^i),
        \label{eq:xstar-ibar}
    \end{align}
    where in the second equality I used equation~\ref{eq:xbar-ibar}, and the fact that, by Lemma~\ref{lemma:ibar-fillslast}, $x^T_k=0$ for all $v_k \geq v_{\bar{i}}$. Finally, if I multiply and divide both sides of equation~\ref{eq:xstar-ibar} by $v_{\bar{i}}$ and $v_{\underline{i}}$, respectively, and I combine the resulting equation with equation~\ref{eq:weight} I get the desired result:
    \[
    \sum_{j=0}^{T-1}y\left(x^{S_j}\right)=\frac{v_{\bar{i}}\bar{x}^*_{\bar{i}}}{v_{\underline{i}}}+1-\bar{x}^*_{\underline{i}}=\underline{\lambda}.
    \]
\end{proof}

\subsection{Step 3}
\begin{proof}[Only if direction of Theorem~\ref{theorem:characterization-WC}]
    Suppose that $\underline{u}(\sigma,f)\geq u^*$ for all $\sigma \in \Sigma(f)$. Towards a contradiction, assume that $f_s \leq \underline{\lambda}$ for some label $s$. Let $T \geq 0$ be the smallest step at which the mass of consumers assigned to all extreme markets up to $T$, $\left\{x^{S_j}\right\}_{j=0}^T$, is greater or equal than $f_s$. That is, $\sum_{j=0}^{T-1}y\left(x^{S_j}\right) < f_s \leq\sum_{j=0}^{T} y\left(x^{S_j}\right)$ (with the convention that the empty sum is equal to 0). Then, define $\sigma(\cdot \lvert s)$ as:
    \[
    \sigma(\cdot \lvert s)=\sum_{j=0}^{T-1}\frac{y\left(x^{S_j}\right)}{f_s}x^{S_j}+\left(1-\sum_{j=0}^{T-1}\frac{y\left(x^{S_j}\right)}{f_s}\right)x^{S_T}.
    \]
    
    See that $\sigma(\cdot \lvert s)$ is a weighted average of all extreme markets up to step $T$. Therefore, $\sigma(\cdot \lvert s)$ is a well defined market; i.e., $\sigma(\cdot \lvert s) \in \Delta(V)$. Furthermore, since $f_s \leq \underline{\lambda}$, Lemma~\ref{lemma:weight=lambda} implies that $v_{\bar{i}} \in \supp x^{S_j}$ for all $j \leq T$. Therefore, $v_{\bar{i}}$ is an optimal price for all extreme markets up to step $T$, and thus $v_{\bar{i}}$ is an optimal price for segment $s$. Moreover, because the monopolist breaks ties against consumers, he must set a price weakly greater than $v_{\bar{i}}$ for segment $s$. Additionally, equation \ref{eq:segmentation} ensures that $\sigma(\cdot \lvert s)f_s \leq x^*$, with strict inequality for some $v_k \in V$. Hence, I am not assigning more mass of consumers than the total available for any valuation.

    For any other label $s' \neq s$ set $\sigma(\cdot \lvert s')$ equal to:
    \[
    \sigma(\cdot \lvert s')=\frac{x^*-\sigma(\cdot \lvert s)f_s}{1-f_s}.
    \]
    
    In other words, I am distributing evenly the remaining mass of consumers of each valuation across all labels different from $s$. Note that $\sigma$ is a feasible segmentation given the marginals $x^*$ and $f^*$. Moreover, the monopolist must set the same price for every segment $s' \neq s$, which is no lower than $v_{i^*}$. Indeed, suppose that the profits obtained from setting a price equal to $v_j$ are higher than charging $v_{i^*}$ and $v_j < v_{i^*}$. Then,  $v_j \in \supp \sigma(\cdot \lvert s')$, which is equivalent to $x^*_j > \sigma(j \lvert s)f_s$; thus, $v_j \in \supp x^{S_j}$ for all $j \leq T$. However, this implies that $v_j$ is also an optimal price in segment $s$. Hence,  $v_j$ leads to higher profits than $v_{i^*}$ for the aggregate market, a contradiction.

    Since the consumers in segment $s$ pay a price higher than $v_{i^*}$ without counterbalancing benefits to consumers in the remaining segments, $\sigma$ leads to a consumer surplus lower than $u^*$ but this is a contradiction. 
\end{proof}

\section{Proofs of results in Section~\ref{sec:results}}
\label{sec:proofs-UD}

\begin{proof}[If direction of Theorem~\ref{theorem:characterization-UD}]
    Suppose that $f \in WC$ and $f_s<\bar{\lambda}$ for some label $s$. Then, define $\sigma(\cdot \lvert s)$ as follows. Let $\sigma(j \lvert s)=\frac{x^*_j}{f_s}$ for all $v_j < v_{i^*}$, and for all valuations $v_j \geq v_{i^*}$ set $\sigma(j \lvert s)$ such that two conditions are satisfied: 
    \begin{enumerate}[label=(\roman*)]
    \item $\sigma(j \lvert s) \leq \frac{x^*_j}{f_s}$ and 
    \item $\sum_{j=i^*}^{K}\sigma(j \lvert s)=\frac{\bar{x}^*_{i^*}}{f_s}+1-\frac{1}{f_s}$.
    \end{enumerate}

    Some observations of the above segment. First, it is always possible to find numbers that satisfy conditions (i) and (ii). Indeed, if I set $\sigma(j \lvert s)=\frac{x^*_j}{f_s}$ for all $v_j \geq v_{i^*}$ I will get that $\sum_{i^*}^{K}\sigma(j \lvert s)=\frac{\bar{x}^*_{i^*}}{f_s}$, which is greater than $\frac{\bar{x}^*_{i^*}}{f_s}+1-\frac{1}{f_s}$, then I can reduce $\sigma(j \lvert s)$ for any $v_j$ until both terms are equal. Second, I am assigning the whole mass of consumers with a valuation lower than $v_{i^*}$, but I am not assigning the whole mass of consumers with a valuation weakly higher than $v_{i^*}$. Third, by construction $\sum_{j=1}^{K}\sigma(j \lvert s)=1$, however this does not imply that $\sigma(\cdot \lvert s) \in \Delta(V)$, as it is possible that $\sum_{j=1}^{i^*-1}\frac{x^*_j}{f_s}>1$. For now, assume that $\sum_{j=1}^{i^*-1}\frac{x^*_j}{f_s}<1$ such that $\sigma(\cdot \lvert s) \in \Delta(V)$.

    Next, I want to show that the monopolist will charge a price lower than $v_{i^*}$ in segment $s$. Since $f \in WC$, Corollary~\ref{coro:prices} implies that, for all feasible segmentations, the price in every segment must be weakly lower than $v_{i^*}$. Then, it is enough to check that there exists some $v_k < v_{i^*}$ such that the profits from charging $v_k$ are higher than the profits from charging $v_{i^*}$. The profits from charging $v_k$ and $v_{i^*}$ are equal to:
    \begin{align}
        v_k\sum_{j=k}^{K}\sigma(j \lvert s)&=v_k\left(\sum_{j=k}^{i^*-1}\sigma(j \lvert s)+\sum_{j=i^*}^{K}\sigma(j \lvert s)\right) \nonumber \\
        &=v_k\left(\sum_{j=k}^{i^*-1}\frac{x^*_j}{f_s}+\frac{\bar{x}^*_{i^*}}{f_s}+1-\frac{1}{f_s}\right) \nonumber \\
        &=v_k\left(\frac{\bar{x}^*_{k}}{f_s}+1-\frac{1}{f_s}\right), \label{eq:profits-vk}\\
        v_{i^*}\sum_{j=i^*}^{K}\sigma(j \lvert s)&=v_{i^*}\left(\frac{\bar{x}^*_{i^*}}{f_s}+1-\frac{1}{f_s}\right). \label{eq:profits-istar}
    \end{align}
    Hence, it is enough to check that~\ref{eq:profits-vk} is greater than \ref{eq:profits-istar} for some $v_k < v_{i^*}$:
    \begin{align}
        v_k\left(\frac{\bar{x}^*_{k}}{f_s}+1-\frac{1}{f_s}\right) &> v_{i^*}\left(\frac{\bar{x}^*_{i^*}}{f_s}+1-\frac{1}{f_s}\right) \nonumber\\
        \frac{v_{i^*}-v_{k}-(v_{i^*}\bar{x}^*_{i^*}-v_{k}\bar{x}^*_{k})}{f_s} &> v_{i^*}-v_{k} \nonumber \\
        1-\frac{v_{i^*}\bar{x}^*_{i^*}-v_{k}\bar{x}^*_{k}}{v_{i^*}-v_{k}} &> f_s.\label{eq:upbound-f-UD-proof}
    \end{align}
    Inequality~\ref{eq:upbound-f-UD-proof} holds for some $v_k < v_{i^*}$, since $f_s < \bar{\lambda}=\max_{v_j < v_{i^*}} \; 1-\frac{v_{i^*}\bar{x}_{i*}-v_j\bar{x}^*_j}{v_{i^*}-v_j}$. Therefore, the monopolist charges a price lower than $v_{i^*}$. It remains to define $\sigma$ for the other labels $s' \neq s$. For this purpose, let $\left\{\sigma(\cdot \lvert s')\right\}_{s \neq s'}$ be any collection of distributions that together with $\sigma(\cdot \lvert s)$ is consistent with $f$ and $x^*$. These distributions exist because, for segment $s$, I am not assigning more mass of consumers than the total available for any valuation. Moreover, Corollary~\ref{coro:prices} implies that the price charged by the monopolist in every segment $s' \neq s$ is weakly lower than $v_{i^*}$. Then, all consumers with a valuation weakly higher than $v_{i^*}$ pay a price weakly lower than $v_{i^*}$, and, since I assumed that $\sum_{j=1}^{i^*-1}\frac{x^*_j}{f_s}<1$, a positive mass of them pay a price strictly lower than $v_{i^*}$. Thus, consumer surplus increases relative to $u^*$.

    If $\sum_{j=1}^{i^*-1}\frac{x^*_j}{f_s} \geq 1$ the result still holds. Particularly, I can now construct a $\sigma(\cdot \lvert s)$ such that for a sufficiently small $\varepsilon \in (0,1)$ the following 3 conditions are satisfied: 
    \begin{enumerate}[label=(\roman*)]
    \item $\sigma(j \lvert s) \leq \frac{x^*_j}{f_s}$ for all $v_j \in V$, 
    \item $\sum_{j=1}^{i^*-1}\sigma(j \lvert s)=1-\varepsilon$ and $\sum_{j=i^*}^{K}\sigma(j \lvert s)=\varepsilon$, and 
    \item the monopolist charges a price lower than $v_{i^*}$ in segment $s$.
    \end{enumerate}
    Then, using the same logic as above, one will get that for the other segments, the price must be weakly lower than  $v_{i^*}$; therefore, consumer surplus increases relative to $u^*$.
\end{proof}
\end{appendix}

\newpage
\bibliography{references.bib}
\bibliographystyle{aer}

\end{document}